\documentclass[superscriptaddress,nofootinbib,10pt]{revtex4}
\usepackage{graphicx,amssymb,amsmath,slashed}

\newcommand{\beq}{\begin{equation}}
\newcommand{\eeq}{\end{equation}}

\newcommand{\im}{\mathrm{Im}}

\begin{document}

{\tiny CERN-PH-TH/2012-031}

\title{\bf  Charming CP Violation and Dipole Operators from RS Flavor Anarchy}

\author{C\'edric Delaunay}
\affiliation{CERN, Theoretical Physics, CH-1211 Geneva 23,
Switzerland}

\author{Jernej F.\ Kamenik}
\affiliation{J. Stefan Institute, Jamova 39, P. O. Box 3000,
1001 Ljubljana, Slovenia}
\affiliation{Department of Physics, University of Ljubljana, Jadranska 19, 1000 Ljubljana, Slovenia}

\author{Gilad Perez}
\affiliation{CERN, Theoretical Physics, CH-1211 Geneva 23, Switzerland}
\affiliation{Department of Particle Physics and
Astrophysics, Weizmann Institute of Science, Rehovot 76100,
Israel}

\author{Lisa Randall}
\affiliation{\mbox{Department of Physics, Harvard University, Cambridge, MA 02138}}

\begin{abstract}
Recently the LHCb collaboration reported evidence for direct CP violation in charm decays. The value is  sufficiently large  that  either substantially enhanced  Standard Model contributions  or non-Standard Model physics is required to explain it. In the latter case only a limited number of possibilities would be consistent with other existing flavor-changing constraints.
We show that warped extra dimensional models that explain the quark spectrum through flavor anarchy can naturally give rise to contributions of the size required to explain the the LHCb result. The $D$ meson asymmetry arises through a sizable CP-violating contribution to  a chromomagnetic dipole operator.  This happens naturally without introducing inconsistencies with existing constraints in the up quark sector.
 We discuss some subtleties in the loop calculation that are similar to those in Higgs to $\gamma \gamma$.
Loop-induced dipole operators in warped scenarios and their composite analogs exhibit non-trivial dependence on the Higgs profile, with the contributions monotonically decreasing when the Higgs is pushed away from the IR brane.
We show that the size of the dipole operator quickly saturates as the Higgs profile approaches the IR brane, implying small dependence on the precise details of the Higgs profile when it is quasi IR localized. We also explain why the calculation of the coefficient of the lowest dimension 5D operator is guaranteed to be finite.
This is true not only in the charm sector but also with other radiative processes such as electric dipole moments, $b\to s\gamma$, $\epsilon'/\epsilon_K$ and $\mu\to e\gamma$. We furthermore discuss the interpretation of this contribution within the framework of partial compositeness in four dimensions
and highlight some qualitative differences between the generic result of composite models and that obtained for dynamics that reproduces the warped scenario.
\end{abstract}

\maketitle

%%%%%%%%%%%%%%%%%%%%%%%%%%%%%%%%%%%%%%%%%%%%%%%%%%%%%%%%%%%
\section{Introduction}
%%%%%%%%%%%%%%%%%%%%%%%%%%%%%%%%%%%%%%%%%%%%%%%%%%%%%%%%%%%
Recently the LHCb collaboration reported  $3.5\sigma$ evidence for a non-zero
value of the difference between the time-integrated CP asymmetries in the
decays $D^0 \to K^+K^-$ and $D^0 \to \pi^+\pi^-$~\cite{lhcb} $\Delta a_{CP} \equiv a_{K^+K^-} - a_{\pi^+\pi^-}$, where
\beq
a_f \equiv \frac{\Gamma(D^0 \to f) - \Gamma (\bar D^0\to f)}{\Gamma(D^0 \to f) + \Gamma (\bar D^0\to f)}\,.
\eeq
Combined with other measurements of these CP asymmetries~\cite{CDF10784, Aaltonen:2011se, Staric:2008rx, Aubert:2007if,HFAG}, the present world average is
\begin{equation}
\Delta a_{CP} = -(0.67\pm 0.16)\%\,.
\label{eq:acpExp}
\end{equation}

The effective weak Hamiltonian relevant for hadronic
singly-Cabibbo-suppressed $D$ decays renormalized at a scale $m_c < \mu< m_b$ is
  \begin{equation}\label{SM}
\mathcal H^{\rm SM}_{|\Delta c| = 1} =
   \frac{G_F}{\sqrt 2}
  \sum_{q=s,d}\lambda_q\sum_{i=1,2}  C^q_i Q_i^q + {\rm h.c.} + \ldots\,,
\end{equation}
where $\lambda_q = V_{cq}^*V_{uq}$,
$Q^q_1 = (\bar u  q)_{V-A}\, (\bar q c)_{V-A}\,,$
$Q^q_2 = (\bar u_\alpha  q_\beta)_{V-A}\, (\bar q_\beta  c_\alpha)_{V-A}\,,$
and $\alpha,\beta$ are color indices. Dots denote neglected Standard Model (SM) penguin operators with tiny Wilson coefficients.
 In the SM, as well as within its minimally flavor violating extensions~\cite{MFV}, contributions of the Hamiltonian $\mathcal H^{\rm SM}_{|\Delta c| = 1}$ to $\Delta a_{CP}$ are suppressed relative to the leading CKM terms factored out in Eq.~\eqref{SM} by $|V_{cb}V_{ub}| /
|V_{cs}V_{us}| \approx 0.07\%$ and are therefore expected to be small~\cite{Grossman:2006jg}.
 However, since the charm
scale is not far from $\Lambda_{\rm QCD}$, non-perturbative enhancements leading
to substantially larger values cannot be excluded~\cite{Golden:1989qx} (see also~\cite{Brod:2011re}).

Nonetheless, without a substantial enhancement, the SM contribution would be too small to explain  current observations.
Moreover extensions of the SM generally also have difficulty accommodating the measured value without conflicting with existing stringent flavor-changing constraints \cite{arXiv:1111.4987}, since loop
effects can produce other flavor violations in excess of their experimental values. This however is not true for the chromomagnetic dipole operators~\cite{Grossman:2006jg,Giudice:2012qq} due to the light quark mass suppression that is essential given the operators' helicity structure.
Any new physics contribution to these operators can be encoded in
\begin{eqnarray}
\mathcal H^{\rm chromo}_{|\Delta c| = 1} &=& \frac{G_F}{\sqrt 2}
   (C_{8}  Q_{8} + C'_8 Q^{\prime}_8)+ {\rm h.c.}\,,
\label{eq:HNP}
\end{eqnarray}
where
$Q_{8}=-g_s\, m_c\, \bar u \sigma_{\mu\nu}
  (1+\gamma_5) T^a G_a^{\mu\nu} c$ and
%$Q'_{8} = - g_s\, m_c\, \bar u \sigma_{\mu\nu}
%  (1-\gamma_5) T^a G_a^{\mu\nu} c\,$.%\label{Ops8} $
$Q'_8$ obtained from $Q_8$ with $\gamma_5\to-\gamma_5$.
 The contributions of such operators to
$\Delta a_{CP}$ are given by~\cite{arXiv:1111.4987}
\beq
 \label{delta_acp}
\Delta a^{\rm chromo}_{CP} \approx 9\, \im
(R^K_8  + R^{\pi} _8) \sum_{i = 8,8'} \im (C_i)  \,,
\eeq
where $R^{P} _8$ denote the relevant operator
hadronic matrix element ratios. Following~\cite{Grossman:2006jg} we find
\beq
|R^P_8| = \frac{16\pi}{3|C_1|} \alpha_s \left[ \frac{1+r^P_\chi/3}{1+C_2/3C_1}  \right] \simeq 12(9) ~{\rm for}~ P=K(\pi)\,,
\label{eq:R8}
\eeq
where we have used $\alpha_s(m_c)\simeq 0.35$, $C_2(m_c) \simeq -0.4$, $C_1(m_c) \simeq 1.2$~\cite{Buchalla:1995vs}, and $r^P_\chi = 2m_P^2/m_c(m_q+m_u)\simeq 3.6 (1.7)$ for $P=K(\pi)$ and $q=s(d)$ using inputs from~\cite{PDG}.
 We note that under the above normalization convention for the dipole operators in Eq.~\eqref{eq:HNP}
and using naive factorization, their matrix elements are enhanced by a factor of ${\cal O}$(10) relative to the tree and penguin operators. Thus, given $\mathcal O(1)$ Wilson coefficients of the dipole operators in Eq.~\eqref{eq:HNP}, the scale required to saturate the experimental value in Eq.~(\ref{eq:acpExp}) is
 $\Lambda_8\sim 20\,{\rm TeV}\,$.
This is comparable to the scale $\Lambda_{4\rm f}\sim 10{\rm\, TeV}$ that was found for the four fermion operators in~\cite{arXiv:1111.4987}.
Notice that for this calculation we have evolved the Wilson coefficients computed at the  scale $\Lambda_8$ down to the charm mass scale. The leading order anomalous dimension for $Q^{(\prime)}_8$ is $\hat \gamma = 28/3$ in the conventions of~\cite{Buchalla:1995vs}.  Consequently, the
Wilson coefficients at the $m_c$ scale are
%\beq
$C^{(\prime)}_8(m_c) = \eta  C^{(\prime)}_8\big(\Lambda_8\big)\,$,
%\eeq
where
\beq \label{eta}
\eta = \left[\frac{\alpha_s(m_b)}{\alpha_s(m_c)}\right]^{14/25}  \left[\frac{\alpha_s(m_t)}{\alpha_s(m_b)}\right]^{14/23} \left[\frac{\alpha_s(\Lambda_8)}{\alpha_s(m_t)}\right]^{2/3}\,.
\eeq
\\

%%%%%%%%%%%%%%%%%%%%%%%%%%%%%%%%%%%%%%%%%%%%%%%%%%%%%%%%%%%
\section{Dipole Operator in RS}
%%%%%%%%%%%%%%%%%%%%%%%%%%%%%%%%%%%%%%%%%%%%%%%%%%%%%%%%%%%
One important feature of the bulk Randall-Sundrum (RS) framework~\cite{RS} is that beyond explaining the electroweak scale,
 it can lead to hierarchies
in quark and lepton masses without any hierarchies in fundamental flavor parameters~\cite{RS,GW}.
This is achieved when different flavors are localized in separate regions along the extra dimension~\cite{GN,RSoriginal} through their different bulk masses. In this setup, gauge bosons, fermions, and the Higgs boson are in the bulk with light quarks localized in the UV while the Higgs and KK modes are localized in the IR. The assumption is that mixing angles arise (up to order one factors from the Yukawas) as the ratios of left-handed wave functions while the ratios of masses are determined by the ratios of right handed wave functions (divided by mixing angles).
The interesting feature in the flavor sector is that although an explicit GIM mechanism is absent, flavor-changing processes are suppressed by wavefunctions related to quark masses and mixing angles and are therefore reasonably consistent with observations~\cite{aps}.
However, some flavor changing processes can exceed their SM values, such as $\Delta a_{CP}$ which we consider here (for observables related to the down sector see~\cite{FirstepsK,Agashe:2008uz,GIP}).
We now show that the version of RS that accounts for quark and lepton masses
through anarchic up-type Yukawa matrices can give a sufficiently large contribution to $\Delta a_{CP}$ through the
chromomagnetic operators in Eq.~\eqref{eq:HNP}.
 As with any model contributing through the chromomagnetic operator, the contribution can in principle be large enough to account for $\Delta a_{CP}$ without introducing overly large flavor-changing effects in other $D$ meson processes~\cite{Grossman:2006jg,Giudice:2012qq}. The important additional feature in RS flavor models is that all flavor-changing processes are suppressed by wavefunctions, with no additional requirement of near-degeneracy in masses, as there would be for example in the squark sector of a supersymmetric model. Such a requirement would be difficult to accommodate consistently with the requisite off-diagonal flavor entries in a model with only symmetries as a constraint.

We work in a slice of AdS$_5$ spacetime
whose fifth (conformal) coordinate $z$ is bounded by two
branes at $z_{UV}%=M_{\rm \overline{Pl}}^{-1}
\sim
(10^{19}\,$GeV$)^{-1}$ in the UV and $\bar{z}\sim\,$TeV$^{-1}$ in
the IR. %, where $M_{\rm \overline{Pl}}$ is the reduced Planck mass.
We also assume that the
Higgs field, $H$, is a bulk field with vacuum expectation value
(VEV)
\beq\label{Hprofile}
\langle H\rangle= v\bar{z}/z_{UV}^{3/2}
\sqrt{1+\beta}(z/\bar{z})^{2+\beta}
\eeq
with $v\simeq 246\,$GeV. $\beta$
parameterizes the localization of the Higgs VEV in the bulk. We assume for simplicity that the SM-like Higgs fluctuation $h$  has the same profile as its VEV along the fifth dimension, which is a good approximation up to $\mathcal{O}(m_h^2/m_{\rm KK}^2)$ where $m_{\rm KK}$ is the KK scale.  The case $\beta=0$
corresponds to gauge-Higgs unified models (see {\it e.g.}~\cite{GHU}).

\begin{figure}[t]
\begin{center}
\includegraphics[width=0.4\textwidth]{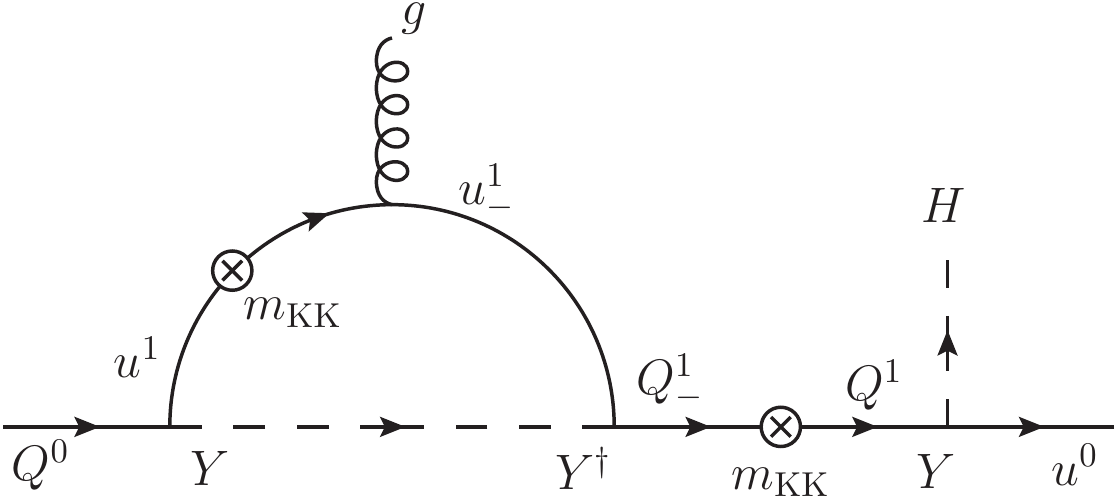}
\caption{{Leading one-loop diagram that contributes to the operator $Q_8$ from neutral Higgs exchange. Arrows denote the flow of charge through the diagram. Diagrams with a gluon radiated from the external quark legs are ignored as they contribute only to the gauge coupling renormalization.}}
\label{fig:RSdiags}
\end{center}
\end{figure}

Since in RS the only terms that generate interactions between SU(2) doublets and singlets
are the 5D Yukawa interactions, the Hamiltonian in Eq.~\eqref{eq:HNP} is induced  through a one-loop contribution involving exchange of
KK fermions and the Higgs boson~\cite{aps}. Some of the relevant diagrams are shown in Fig.~\ref{fig:RSdiags}.
Following Refs.~\cite{Agashe:2008uz,GIP}, an explicit evaluation of the one-loop amplitude yields
\beq
C_8(m_{\rm KK})=U_L^{12}\frac{\sqrt{2}Y_5^2}{16\pi^2 G_F m_{\rm KK}^2}\mathcal{O}_\beta\,,
\label{eq:RSdipole}
\eeq
where $m_{\rm KK}\simeq 2.45/R'$ is the KK scale and $Y_5$ is the 5D Yukawa coupling in appropriate units
of the AdS curvature. $C_8'$ is obtained through the replacement $U_L\to U_R$ and $U_{L(R)}^{12}$ denotes the 1-2 mixing angle in the left- (right-) handed up quark sector.
Under the assumption of anarchic RS flavor parameters, the mixing angles are $U_L^{12}\sim f_{Q^1}/f_{Q^2} \simeq \lambda_c$ and $U_R^{12}\sim f_u/f_c \simeq (m_u/m_c)/\lambda_c$, where $\lambda_c\simeq 0.23$ is the sine of the Cabibbo angle, and $C_8^{(\prime)}$ carries an arbitrary $\mathcal{O}(1)$ CPV phase provided by the anarchic 5D Yukawa.
 $f_{x^i}$
 is the value of the $x^i$ quark zero-mode profile on the IR brane. The contribution to the chromomagnetic operator involving the left-handed up quark, $Q_{8}$, can be large in the RS scenario because of the sizable ratio of the first two generation wave functions, which is comparable to the relatively large Cabibbo angle.

  The function $\mathcal{O}_\beta$ parameterizes the Higgs profile overlap with the KK state wavefunctions, and depends significantly on the Higgs field localization parameter $\beta$ as we describe below.
Plugging Eq.~\eqref{eq:RSdipole} and Eq.~\eqref{eq:R8} back into Eq.~\eqref{delta_acp}, along with RS parameters yielding the correct quark masses~(see {\it e.g.}~\cite{Isidori:2010kg}) and including a running factor from $3\,{\rm TeV}$ to the charm scale of $\eta \simeq0.4$, we find
\beq
\left|\Delta a_{CP}^{\rm chromo}\right|_{\,\rm RS}\simeq 0.6\% \times  \left(\frac{\mathcal{O}_{\beta}}{0.1}\right) \,\left(\frac{Y_5}{6}\right)^2\left(\frac{3\,{\rm TeV}}{m_{\rm KK}}\right)^2\,,
\eeq
which is of the right size required by Eq.~\eqref{eq:acpExp}. For the sake of definiteness we have assumed a maximally delocalized Higgs ($\beta=0$), as arises in gauge-Higgs unification models. In this case the overlap function  is $\mathcal{O}_\beta\simeq 0.1$ for UV-localized first two generations.

As we already mentioned, $\mathcal{O}_\beta$ has a strong dependence on the Higgs profile. In particular as we show below  for large values of $\beta\gtrsim 10$ mimicking a Higgs profile very peaked towards the IR brane the overlap becomes $\mathcal{O}_\beta\sim \mathcal{O}(1)$. The LHCb result is thus reproduced with a smaller 5D Yukawa of $Y_5\simeq 2$, similarly to  generic composite models~\cite{KerenZur:2012fr}.
We study below the dependence of $\mathcal{O}_\beta$ on the Higgs field localization in the bulk and we present the implications of $\Delta a_{CP}$ within up flavor anarchic RS models as a function of the Higgs overlap function.

%%%%%%%%%%%%%%%%%%%%%%%%%%%%%%%%%%%%%%%%%%%%%%%%%%%%%%%%%%
\subsection{Higgs profile dependence}

We now evaluate the RS contributions to the dipole operator for varying Higgs profiles. We describe some subtle features of the calculation, including the Higgs profile and cutoff dependence, and, in particular,  the result with a properly regularized brane-localized Higgs. We address issues related to cutoff dependence, where we need to account for both the UV cutoff and Higgs regularization (for the brane-localized case) as discussed in~\cite{HY}. We furthermore address the critical dependence on the Higgs coupling to the ``wrong chirality modes", the KK modes with opposite chirality to the zero-modes.  We find the wrong-chirality couplings and the heavy KK modes near the cutoff play a critical role and furthermore one has to account for the order of removing the cutoff of the 5D effective theory and the Higgs width parameter. We describe here our computation method and some qualitative features of the result, which we develop in detail in the Appendices. We show a simple argument for the necessity of both the wrong-chirality Higgs coupling (which in turn helps explain the finiteness of the result) and the critical contribution of heavy KK modes (those with KK number inversely proportional to the Higgs profile width) in the mass insertion perturbative approach.

As in  Ref.~\cite{Azatov:2009na}, we assume the Higgs is a bulk field whose localization is controlled by the free parameter $\beta$ defined in Eq.~\eqref{Hprofile}. The larger $\beta$, the more the Higgs profile is peaked towards the IR brane. We evaluate the relevant dipole contributions at one-loop using KK decomposition and sum the fermion KK towers up to arbitrarily high KK masses, thus capturing any possible sensitivity to high mass modes. Besides determining the dependence of the dipole contributions to the Higgs profile in the bulk, such a calculation can also serve as a regularization procedure for the IR localized Higgs case~\cite{Azatov:2009na} by taking the formal limit $\beta\to \infty$.

For any finite $\beta$  the KK sum is guaranteed to converge
  since the 5D operator $Y^3_5g_5 \bar{Q}\sigma_{\mu\nu}G^{\mu\nu}Hu$ is finite for a bulk Higgs~\cite{aps}. On the other hand naive 5D power counting suggest that the same operator might be logarithmically UV divergent when the Higgs is an IR localized field and therefore its dipole contribution,  being UV sensitive,  would be incalculable. The difference can be accounted for by the loss of momentum conservation for a brane-localized Higgs, which would allow two independent KK sums for fermions in the loop, thereby increasing the degree of divergence.

 A closer look at the bulk equations of motion shows that not all the fermion wave-functions are unambiguously defined for a delta-function Higgs. In any case, such a Higgs profile ceases to make sense for a regulated brane with finite thickness of order of the cutoff. Clearly the Higgs profile will not be more localized than the brane thickness. The IR brane Higgs can however be regularized by spreading its profile on the brane or even into the bulk~\cite{Azatov:2009na}. A Higgs profile like in Eq.~\eqref{Hprofile} with a large but finite $\beta$ would provide such a regulator. Using the procedure outlined above and taking the large $\beta$ limit we find that the dipole contributions converge to a finite value, from which we conclude that the latter are one-loop finite and thus calculable even for an IR localized Higgs.\footnote{We note that the same regularization procedure is expected to render a finite Higgs-gluons couplings as with a bulk Higgs they arise from the $|Y_5^2| g_5^2 H^\dagger HG^{\mu\nu} G_{\mu\nu}$ operator, which is irrelevant in 5D.} We now present a simple argument to justify this result.

The first important point in understanding the result is the essential insertion of the Higgs coupling to the wrong chirality modes. This result does not arise from a symmetry of the theory but rather follows from the particular structure the Yukawa coupling matrix has in the limit where the wrong chirality Yukawa is absent.

We show that the 1PI contributions are controlled by the Yukawa interaction coupling of the heavy fermion of wrong chirality, opposite to that of the chiral states. Notice that a PQ symmetry does account for the need for wrong sign chirality in the case of neutral Higgs exchange. However, this argument does not hold for the charged Higgs, for which both up and down quark couplings to the Higgs field are present. Nonetheless, even when a charged Higgs is exchanged, the relevant diagrams vanish in absence of the wrong chirality Yukawa.

A hint that a structure beyond PQ symmetry is responsible is seen as follows. Although 1PI diagrams with an up-type quark running in the loop are forbidden by the PQ symmetry for a complex Higgs,  diagrams with only the real or imaginary Higgs component in the loop are still formally allowed. The corresponding loop amplitudes have the same form as in Eqs.~\eqref{Ma},\eqref{Mb} and thus the term $\propto y_u^{01}y_u^{11*}y_u^{10}$ vanishes in the $m_H\to 0$ limit for both real and imaginary Higgs components {\it individually}. This observation suggests that in the absence of Yukawa coupling to the wrong chirality states both the neutral and charged Higgs 1PI diagrams vanish even without invoking a PQ symmetry, indicating the PQ symmetry does not suffice to understand the absence of 1PI dipole contributions arising from the $y_u^{11}$ interaction.

We now show the particular structure of the Yukawa coupling matrices   (in the absence of the wrong chirality Higgs coupling) leads to a cancellation of the leading Higgs dependence at each KK level. This structure is not the result of a symmetry of the theory but resembles the mechanism of Nelson-Barr for solving the strong CP-problem~\cite{NelsonBarr}. We focus here on the Yukawa interactions among up-type quarks and the neutral Higgs component $H^0=v+h$, where $v\simeq 246\,$GeV is the Higgs VEV and $h$ is the SM-like Higgs fluctuation. The relevant part of the Lagrangian in Eq.~\eqref{Leff} is
\beq
-\mathcal{L}\supset \bar{u}_L Y_u u_R h + \bar{u}_L M_u u_R +{\rm h.c.}\,,
\eeq
where $u_L=(Q^0,Q^1,u^1_-)$, $u_R=(u^0,Q_-^1,u^1)$ and
\beq
Y_u=\left(\begin{array}{ccc}
y_u^{00} & 0 & y_u^{01}\\
y_u^{10} & 0 & y_u^{11}\\
0 & y_u^- & 0
\end{array}\right)\,,\quad
M_u=\left(\begin{array}{ccc}
y_u^{00}v & 0 & y_u^{01}v\\
y_u^{10}v & m & y_u^{11}v\\
0 & y_u^- v & m
\end{array}\right)\,,
\eeq
where we have assumed $m_Q=m_u=m$. We set now $y_u^-=0$ and show how the one-loop dipole contribution is suppressed in that case. Since $m\gg v$ the two heavy KK states nearly maximally mix through $y_u^{11}$. After diagonalizing the corresponding $2\times2$ block by means of a bi-unitary transformation the heavy eigenmasses are $m_\pm=m \pm y_u^{11}v$.
Below we consider these two states as approximate mass eigenstates and treat their remaining mixing with the zero-mode perturbatively. The projections $y_u^{0\pm}$ and $y_u^{\pm0}$ of the heavy eigenstates onto the $Q^0$ and $u^0$ zero-mode are
\beq
y_u^{0\pm} = \pm \frac{y_u^{01}}{\sqrt2}\left(1\pm \frac{y_u^{11}v}{4m}\right)\,,\quad {\rm and}\quad  y_u^{\pm0} =  \frac{y_u^{10}}{\sqrt2}\left(1\pm \frac{y_u^{11}v}{4m}\right)\,,
\eeq
respectively. Notice the extra relative sign between the two heavy mode projections on $Q^0$, which comes from the fact that one of the two unitary transformations has to involve a diagonal ``phase'' of $\pi$ in order to keep the two eigenmasses positive.  (Alternatively, one could have rearranged the states such that the mass matrix is manifestly positive, in which case a sign explicitly occurs in the mass eigenstate). This sign cancels against the sign of $y_u^{11}$ in the heavy masses  $m_\pm$. For each heavy eigenstate we now show that there is a cancelation at leading order in $v$ in the dipole amplitude between the $y_u^{11}$ correction to the KK mass and the projections on the zero-modes.  The one-loop dipole amplitudes is of the form~\cite{GIP}
\beq
C_g\propto \sum_{j=\pm} \frac{y_u^{0j}y_u^{j0}}{m_j}\,.
\eeq
The leading contribution to the dipole operator contains one chirality flip and is therefore linear in the Higgs VEV $v$. One can extract this linear piece by taking one derivative of the above expression with respect to the Higgs VEV which yields
\beq
v\frac{dC_g}{dv}\big|_{v=0}\propto y_u^{01}y_u^{10}\frac{v}{4m^2} \sum_{j=\pm}\left(y_u^{11}-j \frac{dm_j}{dv}\right)\,.
\eeq
Since $dm_\pm/dv=\pm y_u^{11}$ the leading order contribution  to the dipole operator vanishes for each KK level.

 The above shows that the coupling of the Higgs to the wrong chirality modes is critical to the leading contribution to the dipole operator, and therefore the behavior of the wrong chirality modes near the IR brane plays a crucial role. Since the equations of motion force the wrong chirality fields to vanish (at least in the absence of a Yukawa-dependent delta-function source~\cite{Azatov:2009na}), the result with a delta-function Higgs profile is ambiguous since the delta-function is infinite at the point where those fields vanish. This ambiguity can be resolved by the beta-function regularization mentioned above that gives the Higgs boson a finite thickness in the bulk. This Higgs ``width'' can be taken as small as the brane thickness which must be no greater than the UV cutoff of the theory on the IR brane. The calculation can then be done explicitly with five-dimensional wave-functions in the presence of the nontrivial Higgs profile.

Alternatively, the calculation can be done with perturbative insertions proportional to the Yukawa coupling without solving the full 5D equations of motion. We take the latter approach here and consider the net contribution of KK modes up to the cutoff scale. We will see that as long as the cut off scale is much bigger than the inverse of the width, the result converges to a $\beta$-independent value, but that only heavy KK modes with masses of order the inverse Higgs profile width are relevant.
  At any large but finite $\beta$ the Higgs overlap with fermion KK modes of high enough KK number starts probing the ``bulky'' nature of the Higgs and the
KK sum converges, as dictated by 5D power counting for a bulk Higgs field. The finiteness of the RS contributions for any $\beta$ appears to be consistent with the finding of~\cite{Blanke:2012tv}.

\begin{figure}[t]
\begin{center}
\includegraphics[width=0.4\textwidth]{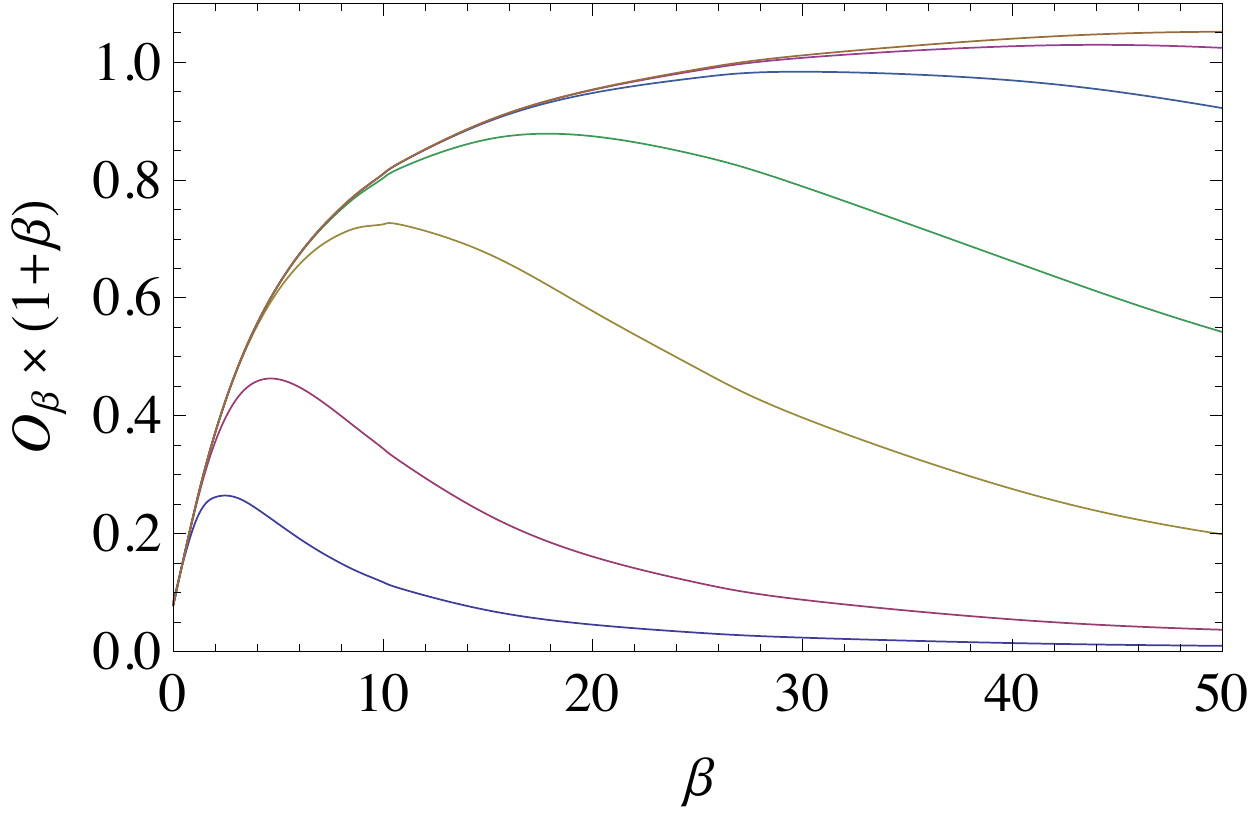}\hspace*{.2cm}
\includegraphics[width=0.4\textwidth]{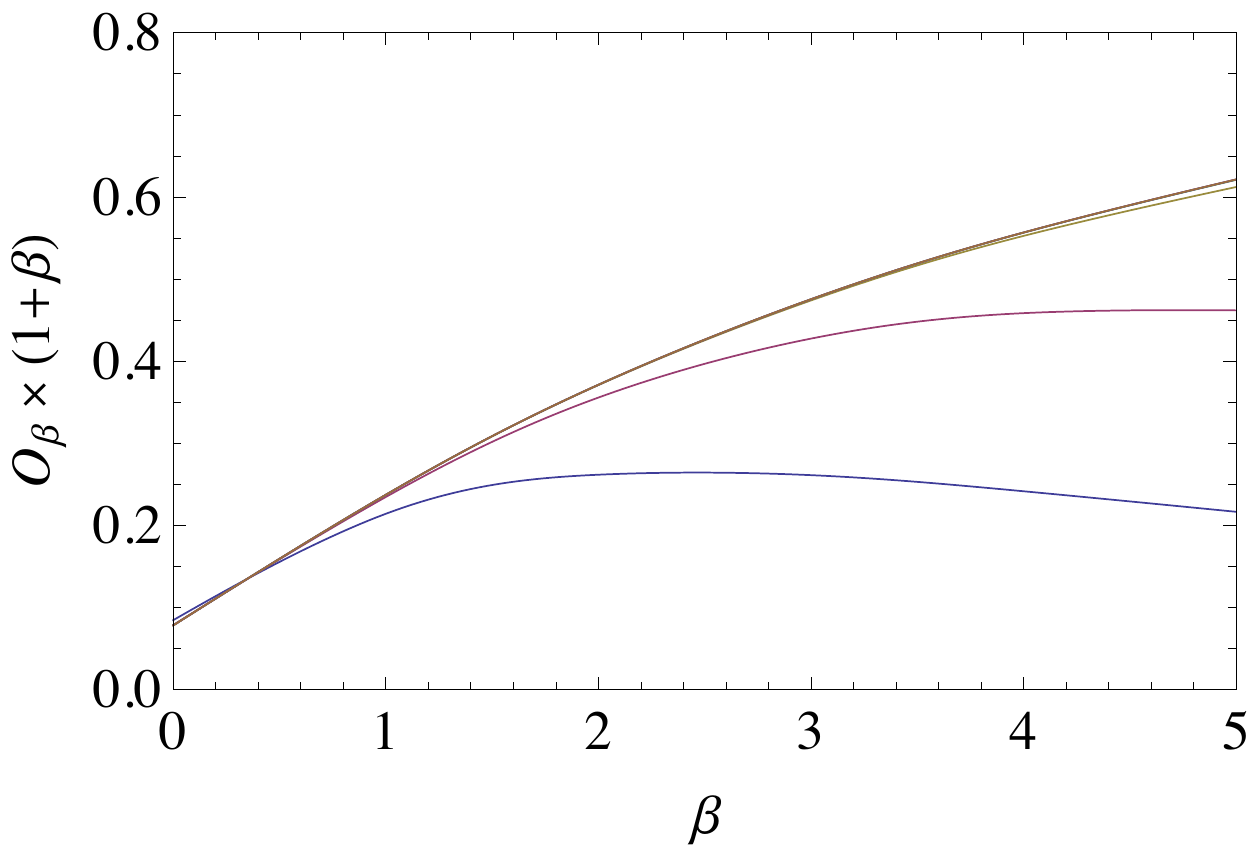}\\
\caption{{Bulk Higgs overlap function charaterizing the size of the dipole contribution at one-loop. We have rescaled the 5D Yukawa by a factor $\sqrt{1+\beta}$ in order to maintain the effective Yukawa finite in the large $\beta$ limit. The various lines show the result of Eq.~\eqref{Obdef} where the KK sums are performed up to 1,2,5,10,20,30,40 KK levels, from bottom to top. For values of $\beta\gtrsim \mathcal{O}(10)$ (left plot) we find $(1+\beta)\mathcal{O}_\beta\sim \mathcal{O}(1)$, the overlap is suppressed for smaller values of $\beta$ (right plot).}}
\label{fig:Obeta}
\end{center}
\end{figure}

The function $\mathcal{O}_\beta$ in Eq.~\eqref{eq:RSdipole} collectively represents the explicit evaluation of the Higgs  overlaps with the KK fermion wave functions as well as the summation over the fermion KK towers. The dominant diagram (shown in Fig.~\ref{fig:RSdiags}) to the dipole amplitude is controlled by the Yukawa coupling to the wrong chiralities~\cite{GIP} as argued above.
The overlap function then parametrically behaves as (see Appendix~\ref{LoopApp})
\beq\label{Obdef}
Y_5^2\mathcal{O}_\beta \equiv \sum_{n,m\geq 1}^{\infty}\frac{y_{0n}(\beta)y_{nm}^{-}(\beta)y_{m0}(\beta)}{y_{00}(\beta)}\times \frac{1}{nm}
\eeq
where $y_{kl}$ corresponds to the effective Yukawa coupling between KK fermions of $k$ and $l$ fermion KK number respectively and $y^{-}$ stands for the effective Yukawa coupling to the wrong chirality KK fermions. The $y_{00}$ factor arises from the SM fermion mass replacement and we have approximated the $n$-th KK level masses by $m_{\rm KK}^{(n)}\simeq nm_{\rm KK}$.
The numerical value of $\mathcal{O}_{\beta}$ as a function of the Higgs localization is shown on the left-hand side plot of Fig.~\ref{fig:Obeta} where the different curves correspond to various numbers of KK modes included in the calculation.
%One clearly sees that the result for large $\beta$ saturates.
Notice that we have rescaled the 5D Yukawa by a factor $\sqrt{1+\beta}$ in order to maintain the effective Yukawa finite (and $\beta$ independent) in the large $\beta$ limit~\cite{Agashe:2008uz}.
As anticipated by~\cite{Azatov:2009na} for large enough $\beta$ the function $\mathcal{O}_\beta$ saturates and only very weakly depends on the  precise value of $\beta$.
Practically the Higgs can be considered a brane-localized field for values as large as $\beta\gtrsim \mathcal{O}(10)$.

We demonstrate this result analytically in Appendix~\ref{flatXD} for a flat extra dimension of size $L$. Besides being simpler to analyze the flat extra dimension case provides a good description of the problem at hand here since the relevant dynamics corresponds to the deep UV regime of the theory and locally, near the IR brane, the details of the bulk geometry should have no practical impact on the result. In particular we show that for a Higgs quasi-localized near the IR brane with a width $\epsilon \ll 1/ Lm_{\rm KK}$  the overlap $\mathcal{O}_\beta$ receives support dominantly
 from a shell of $\sim 1/\epsilon^2$ modes with KK number of $\mathcal{O} (1/\epsilon) $.
 %Note that when $\epsilon/m_{KK}$ is of order the brane width, the modes that contribute have mass of order the cutoff.

A simple argument, in the warped case, for how each KK mode contributes follows from the asymptotic form of the heavy KK modes near the IR brane, where they behave as $\sin(m_n z)$ with $z^{-1}=k e^{k r}$ and $m_n$ is the KK mass of the $n$-th KK level. Let $z=\bar{z} \exp (-\epsilon)$  where we have taken $r-\bar{r}\sim\epsilon/k\sim1/\beta$, with  $\bar{z}=e^{-k \bar{r}}/k$, so that we are still within the region of support of the Higgs wave function. Since $\sin (m_n \bar{z})=0$ by the boundary condition so expanding the exponential and $\sin$ function
we get $\sin (m_n z)=-\cos (m_n \bar{z}) \sin (m_n \bar{z} \epsilon)$. So if $m_n\sim(\epsilon \bar{z})^{-1}$
we are left with the argument of order unity so that the heavy modes are expected to give the dominant contribution.

%with mass of order the cutoff.
  Each KK mode with mass of order $m_{\rm KK}/\epsilon$  contributes with a  $\mathcal{O}(\epsilon^2)$  suppression factor  to the dipole operator. However the sum over all the relevant  modes in the shell yields an $\epsilon$-independent result, thus explaining the saturation observed in Fig.~\ref{fig:Obeta}.
One sees explicitly on Fig.~\ref{fig:Obeta} that for a Higgs width of $1/\beta$ a number of $\mathcal{O}(\beta)$ KK modes needs to be included in order to approach the correct answer. One can also see from the figure that the support for the KK fermions and Higgs overlap separates when the Higgs is peaked further in the IR. When the Higgs is pushed to the bulk, away from the IR brane, ${\cal O}_\beta\ll1$ is expected to have the following behavior: the more the Higgs is pushed onto the bulk, the coupling between the zero modes and the KK fermions becomes more suppressed since it approaches the limit of a flat Higgs profile for which they would be orthogonal.
Furthermore, a bulk Higgs has a larger overlap with the elementary fermions, and the dipole operator is inversely proportional to these couplings. The coupling between the Higgs field and the wrong chirality KK states, that dominates the dipole contribution has non-monotonic but moderate dependence on $\beta$ that is described in more details in Appendix~\ref{flatXD}.
For a maximally delocalized Higgs ($\beta=0$) and UV-localized first two generations
the overlap  amounts to a factor of $\mathcal{O}_{\beta} \simeq 0.1$ while for
 $\beta\gtrsim \mathcal{O}(10)$ we find $\mathcal{O}_{\beta} \simeq \mathcal{O}(1)$. The region of $\mathcal{O}_{\beta}$ for the bulk Higgs case is shown on the right plot of Fig.~\ref{fig:Obeta}, which shows indeed that for low value of $\beta$ the dipole contributions are suppressed and furthermore dominated by the first KK level.
We find the following functional fit to the value of $\mathcal{O}_{\beta}$ that reproduces the numerical result to better than 5\% accuracy over the range $0<\beta <50$
\beq
 \mathcal{O}_{\beta} \simeq F(\beta)= 0.71 + 0.35\,\tanh(0.044\beta)-0.63\exp(-0.27\beta)\,.
\eeq

Notice that  in the perturbative mass-insertion approach we have taken here, the heavy KK modes, with masses of the order of the inverse of the Higgs width, dominate the calculation,  but the dipole operator is nonetheless controlled by a scale of $m_{\rm KK}$ in the final answer.   Although the answer is finite and cutoff-independent, the  dipole contribution could nonetheless be sensitive to higher dimension operators with derivatives in the $z$ direction, since once inserted into the loops we have considered those would yield an unsuppressed amplitude. However, a separation of scales between the inverse width and the cutoff is required to regularize the calculation consistently within the 5D effective theory. Thus as long as there is hierarchy between the inverse width and the cutoff, these higher order operators dependent on the cutoff can be neglected.

\subsection{Implications of $\Delta a_{CP}$ and RS up flavor anarchy}
%%%%%%%%%%%%%%%%%%%%%%%%%%%%%%%%%%%%%%%%%%%%
We find several interesting implications of the observed CP violation in charm decays if it arises from the RS conribution to the chromomagnetic dipole operator. We also outline differences between RS and  generic composite  models in 4D.
%Several comments are in order.
First of all notice that the contribution of $Q'_8$ is negligible in anarchic scenarios due to the additional up-quark mass suppression which leads
to a factor of $(m_u/m_c)/\lambda_c^2\sim 1/40$ suppression relative to the dominant  $Q_8$ contribution.
Second, the central experimental value requires that the 5D Yukawa changes from $\sim 6$ for $\beta=0$ down to
 $\sim2$ for large values of $\beta$.
The $\beta=0$ case requires 5D Yukawa value close to the perturbative bound of $4\pi/\sqrt{N_{\rm KK}}\simeq 7$ for $N_{\rm KK}=3$ perturbative KK states. Such a large Yukawa is further motivated by the need to suppress the contributions to $\epsilon_K$~\cite{Agashe:2008uz}, but is otherwise arbitrary and can potentially lead to large corrections to Higgs production and decay rates~\cite{HY}.
Such a large Yukawa would also enhance the RS contribution to direct CPV in neutral Kaon decays~\cite{GIP}. Both $\epsilon_K$ and $\epsilon'/\epsilon$ constraints can be satisfied with $m_{\rm KK}\gtrsim 5\,$TeV and $Y_5\simeq 5$~\cite{GIP}.
However, clearly such a large KK scale would induce a rather severe little hierarchy problem, so a parametric approach that relaxes the tension with the constraint from down type flavor violation is desired. If the bulk masses are aligned with the down Yukawa matrices,  resulting in a model of up type quark anarchy~\cite{alignment}, the tension is reduced, allowing for a lower KK scale and a large $D$ meson asymmetry.

CP-violating $D$ meson mixing is inversely proportional to the size of the 5D Yukawa in warped models~\cite{Gedalia:2009kh} and it is interesting that the current limit on this effect is saturated for a 5D Yukawa of  $\sim 1.5$ for the brane Higgs case ~\cite{Gedalia:2009kh, Isidori:2010kg}.
With only mild experimental progress the LHCb collaboration would thus start probing the parameter space of  RS with up flavor anarchy and down alignment.
In other words the observation of a CPV signal in $D-\bar D$ mixing could be used to determine the degree of IR localization of the Higgs field, provided that $\Delta a_{CP}$ is indeed coming from the mechanism discussed above.

Note that generic composite models do not capture the distinctive feature of RS models that the composite dynamics is in fact chiral. That is seen by the differing shape of the KK modes for the right chirality and wrong chirality modes, extending the chiral structure of the zero modes. In a deconstructed version of this theory, the extra chiral nature of the RS dynamics means that a two-site model is not sufficient to capture the full composite dynamics. The behavior of the overlap function $\mathcal{O}_\beta$ can be obtained in a model with at least  three sites.   The extra suppression resulting from the overlap correction $\mathcal{O}_\beta$ is inherent to warped models when the Higgs is a bulk field~\cite{Agashe:2008uz,GIP,Csaki:2010aj}. Because the loop factor and the possible overlap suppressions in the RS result in Eq.~\eqref{eq:RSdipole} are not generic to models of four dimensional (4D) composite Higgs with partial compositeness, generic composite models can account for $\Delta a_{CP}$ without as large a Yukawa~\cite{KerenZur:2012fr} which makes these models in principle distinguishable from a bulk Higgs RS model, again through their larger contribution to CP-violating $D$ meson mixing. However, one should keep in mind that generic partial composite models also predict  even larger  down type flavor changing neutral currents which render them somewhat tuned.

%%%%%%%%%%%%%%%%%%%%%%%%%%%%%%%%%%%%%%%%%%%%%%%%%%%%%%%%%%%
 \section{Other observables}
 %%%%%%%%%%%%%%%%%%%%%%%%%%%%%%%%%%%%%%%%%%%%%%%%%%%%%%%%%%%
Given the significant size of the chromomagnetic operator, we explore whether other large dipole contributions can be observable in radiative or rare semileptonic $D$ meson decays. We will see that the RS flavor contributions although sizable, are suppressed compared to the long distance SM contributions that dominate for CP-preserving processes.

The calculation can be encoded in
\begin{eqnarray}
\mathcal H^{\rm EM~dipole}_{|\Delta c| = 1} &=& \frac{G_F}{\sqrt 2}
  (C_{7}  Q_{7} + C'_7 Q^{\prime}_7)+ {\rm h.c.}\,,
\label{eq:HNPgamma}
\end{eqnarray}
with
$Q_{7} = - e\, Q_u m_c\, \bar u \sigma_{\mu\nu}
 (1+\gamma_5) F^{\mu\nu} c\,$, and
%$Q'_{7} = - e\, m_c\, \bar u \sigma_{\mu\nu}
% (1-\gamma_5) F^{\mu\nu} c\,.$
$Q_7'$ obtained from $Q_7$ with $\gamma_5\to -\gamma_5$.
 The contributions of such operators to radiative charm decays can be estimated in the heavy quark expansion for the charm quark. %, where at leading order in QCD and in the static charm quark limit, the $c\to u \gamma$ rate can be written as
%\beq
%\Gamma (c\to u \gamma) = \left(\left| C_7\right|^2 + \left| C'_7\right|^2 \right) \frac{e^2}{\pi} G_F^2 m_c^5\,.
%\Gamma_{c\to u \gamma} = \frac{e^2}{\pi} G_F^2 m_c^5\sum_{i=7,7'}\left| C_i\right|^2\,.
%\eeq
We normalize the radiative rate to the inclusive semileptonic rate $\Gamma(c \to s e^+ \nu_e)$ to suppress charm quark mass dependence. Using the known leading order result for the latter (see e.g.~\cite{Gambino:2010jz}) and inputs from~\cite{PDG} we  estimate the inclusive radiative branching fraction of $D^0 \to X \gamma$ as
%\begin{align}
%Br(D^0 \to X \gamma) & \simeq \frac{\Gamma (c\to u \gamma) }{\Gamma(c \to s e^+ \nu_e)} Br(D^0 \to X e \nu)^{\rm exp} \nonumber\\
\beq
\mathcal B({D^0 \to X \gamma})\simeq\frac{\Gamma({c\to u \gamma}) }{\Gamma({c \to s e^+ \nu_e})} \mathcal B({D^0 \to X e \nu})^{\rm exp}\simeq 5.3 \sum_{i=7,7'}\left| C_i\right|^2,\label{DtoXgamma}
%& \approx 12 \left(\left| C_7\right|^2 + \left| C'_7\right|^2 \right)\,.\label{DtoXgamma}
%\end{align}
\eeq
where at leading order $\Gamma({c\to u \gamma}) = \frac{e^2}{\pi} G_F^2 m_c^5\sum_{i=7,7'}\left| C_i\right|^2\,$.
In RS, one-loop contributions to Eq.~\eqref{eq:HNPgamma} arise not only from diagrams similar to Fig.~\ref{fig:RSdiags} but also from analogues where the photon is emitted from a charged Higgs running in the loop~\cite{Agashe:2008uz}, see Fig.~\ref{fig:EFTphoton}. These yield comparable contributions and we obtain $C_7(m_{\rm KK}) = (5/2) C_8(m_{\rm KK})\,,$ where $C_8(m_{\rm KK})$ is given by Eq.~\eqref{eq:RSdipole}.
%use the former to  estimate  the above branching ratio in RS.
Here again $Q_7$ dominates due to the sizable Cabibbo angle.
%and we find $|C_7(m_{\rm KK})^{\rm RS}| \sim |C_8(m_{\rm KK})|\,,$ where $C_8(m_{\rm KK})$ is given by Eq.~\eqref{eq:RSdipole}.
The pairs of operators $Q_7^{(\prime)}$ and $Q_8^{(\prime)}$ mix under QCD evolution, in particular $C^{(\prime)}_7(m_c) = \eta [\tilde \eta C_7^{(\prime)}(m_{KK}) +8(\tilde \eta-1) C_8^{(\prime)}(m_{KK}) ]$, where $\eta$ is given in Eq.~\eqref{eta} with the replacement $\Lambda_8 \to m_{KK}$ while
\beq\label{tildeta}
\tilde \eta =  \left[\frac{\alpha_s(m_b)}{\alpha_s(m_c)}\right]^{2/25}  \left[\frac{\alpha_s(m_t)}{\alpha_s(m_b)}\right]^{2/23} \left[\frac{\alpha_s(m_{KK})}{\alpha_s(m_t)}\right]^{2/21}\,.
\eeq
Using the values of the above RS parameters to accommodate the $\Delta a_{CP}$ measurement and $\tilde \eta \simeq 0.88$ we obtain from Eq.~\eqref{DtoXgamma}
\beq\label{DtoXgammaRS}
%Br(D^0 \to X \gamma)_{\rm RS}\simeq 4\times 10^{-8}\,,
\mathcal B({D^0 \to X \gamma})^{\rm RS}\simeq 1\times 10^{-8}\,,
\eeq
which is  three orders of magnitude smaller than the estimates of long distance dominated SM contributions to these decays~\cite{LDcug}. Nonetheless, this may be enough to probe such contributions using CP violating asymmetries~\cite{Isidori:2012yx}. In particular for the exclusive $D\to \rho (\omega)\gamma$ final states  maximal RS effects are obtained by marginalizing over the unknown strong phases of the interfering amplitudes, yielding
\beq
|a^{\rm RS}_{\rho (\omega)\gamma}|^{\rm max} \simeq 0.03 \left[ \frac{10^{-5}}{\mathcal B(D^0\to \rho (\omega)\gamma)} \right]^{1/2}\,,
\eeq
an order of magnitude above SM expectations for these observables~\cite{Isidori:2012yx}. Similar effects are also predicted in the $D^0 \to K^+ K^- \gamma$ mode away from the $\phi$ resonance peak.
The present experimental bounds on the two dominant exclusive modes $\mathcal B ({D^0 \to \rho(\omega) \gamma}) < 2.4 \times 10^{-4}$~\cite{PDG} are an order of magnitude above long distance estimates. With the large $D^0$ data samples at the LHCb and the projected Super Flavor Factories, measurements of $a^{}_{\rho (\omega)\gamma}$ at the percent level should be feasible.

The RS effects in  rare semileptonic modes (like $D \to \pi(\rho) \ell^+ \ell^-$ ) are similarly small with respect to long distance contributions. Compared to $c\to u \gamma$,  the estimate is  suppressed by a factor of the fine-structure constant $\alpha$. At leading order in inverse $m_c$
\beq
%\frac{d\Gamma_{c \to u e^+ e^-}}{d\hat s} = \Gamma_{c\to u \gamma} \frac{\alpha}{12\pi} (1-\hat s)^2 \left(1+\frac{2}{\hat s}\right)
%\left(1+ \frac{2\hat m_e^2}{\hat s^2}\right) \sqrt{1-\frac{4\hat m_e^2}{\hat s^2}}  \,,
\frac{d\Gamma({c \to u e^+ e^-})}{d\hat s} = \frac{\alpha\beta_e}{24\pi}(3-\beta_e^2)(1-\hat s)^2 \left(1+\frac{2}{\hat s}\right)\Gamma(c\to u \gamma) \,,
\eeq
where $\hat s = (p_{e^-} + p_{e^+})^2/m_c^2$, $\beta_e=\sqrt{1-4\hat m_e/\hat s}$ and $\hat m_e = m_e/m_c$\,.
%Notice that in the low $\hat s$ region the dominant (long distance) SM contributions are expected to have a similar relationship.
Integrating over $\hat s \in [4 \hat m_e^2, 1]$ we obtain
%\begin{align}
$\mathcal B({D^0 \to X e^+ e^-}) \simeq\,  0.7 \alpha \mathcal B ({D^0 \to X \gamma})\,$.
%\end{align}
Following Eq.~\eqref{DtoXgammaRS},
the RS contribution to the above branching fraction is
\beq\label{DtoXeeRS}
\mathcal B ({D^0 \to X e^+ e^-})^{\rm RS}\simeq 5\times 10^{-12}\,.
\eeq
While the present bounds on the dominant exclusive channels are $\mathcal B({D^0 \to \rho^0 e^+ e^-}) < 10^{-4}$~\cite{PDG}, the long distance contributions again dominate in the SM and are estimated to yield for example $\mathcal B({D^0 \to \rho^0 e^+ e^-})^{\rm SM}\sim 10^{-6}$~\cite{LDcuee}, which is orders of magnitude larger than our RS short distance estimate.
We further note that $O_7^{(\prime)}$ operators do not contribute to purely leptonic $D^0 \to \ell^+ \ell^-$ since the relevant hadronic matrix elements vanish by angular momentum conservation.
%We conclude that within our warped setup there is no implication of the observed direct CPV in charm decays on other radiative D decays.\\

We conclude that the best hope for identifying RS contributions to the CP violating asymmetry in nonleptonic D meson decays might be via CP violating asymmetries in radiative $D$ decays~\cite{Isidori:2012yx} where RS contributions at the percent level are possible. Also crucial is the absence of large CP violation in $D$ meson mixing as we have already discussed.
%, as well as comparable rates in the $\pi \pi $ and $K K$ decay modes.\\

 %%%%%%%%%%%%%%%%%%%%%%%%%%%%%%%%%%%%%%%%%%%%%%%%%%%%%%%%%%%
\section{Conclusions}
%%%%%%%%%%%%%%%%%%%%%%%%%%%%%%%%%%%%%%%%%%%%%%%%%%%%%%%%%%%
To summarize, we have shown that flavor-anarchic warped extra dimension models can generate sizable contributions to $\Delta C=1$ chromomagnetic dipole operators. We have found that with large Yukawa, varying from near-maximal with a maximally delocalized Higgs boson to about a Yukawa of around two  for a brane-localized Higgs, such contributions can be large enough to induce the time-integrated CP asymmetries suggested by the recent $\Delta a_{CP}$ measurements at LHCb and CDF.  The  large Yukawa coupling scenario is rather specific to the RS framework where the dipole operators have additional suppression over generic composite models. We also comment that since the dipole contributions respect SU(3) $(u,d,s)$ flavor symmetry we expect that the individual asymmetries in $D\to\pi\pi$ and $D\to KK$ should be of comparable size, which is a qualitative prediction of this class of models, as are the related asymmetries for other excited states~(for more details see~\cite{Brod:2011re}).

Since the dominant contributions to CP violation in $D-\bar D$ mixing arise from tree level KK gluon exchange and are inversely proportional to the 5D Yukawa~\cite{Gedalia:2009kh,Isidori:2010kg}, they form diagnostics for the nature of the profile of the Higgs or alternatively can tell between generic composite models and the bulk Higgs version of RS as follows: the case where the Higgs is significantly extended into the bulk implies suppressed dipole contribution and a large Yukawa which consequently leads to suppressed contributions to CP violation in $D-\bar D$ mixing. On the other hand, in the brane Higgs limit or generic composite models a smaller Yukawa is required to account for $\Delta a_{CP}$ and the CP violating contributions to $D-\bar D$ mixing could be large, not far from the current experimental bound. Thus, future improvement in the measurement of CP violation in $D$ mixing could therefore in principle shed light on the precise nature of the composite theory.

One should not ignore the fact that composite models suffer from a down quark CP problem (for recent review see~\cite{Isidori:2010kg}). An interesting possibility to
consider is down alignment warped models~\cite{alignment}, which generically circumvent the CP RS Kaon problem but predict up type anarchy. This is perfectly consistent with the current experimental situation and the observed charm CP violation could have been predicted in such a framework well before the measurement.
Furthermore  an indirect support for down alignment and anarchy in the up sector comes from the recent Daya Bay and Reno results~\cite{An:2012eh}, which reported a relatively large value for the only (thus far) unknown neutrino mixing angle. A reasoning similar to that of the quark sector leads to a successful model of leptons in the context of neutrino flavor anarchy and furthermore requires that all mixing angles are reasonably large~\cite{Perez:2008ee}.

\acknowledgments
We thank Kaustubh Agashe, Alexanedr, Azatov, Csaba Csaki, Gian Giudice,  Christophe Grojean, Ulrich Haisch, Gino Isidori, Zoltan Ligeti, Yasinori Nomura, Michele Papucci, Riccardo Rattazzi, Michele Redi,  Sheldon Stone, Philip Tanedo, Yuhsin Tsai, Andi Weiler and Guy Wilkinson for useful discussions.
LR also thanks the theory division of CERN and UC Boulder for their hospitality when part of this work was done.
The work of JFK was supported in part  by the Slovenian Research Agency.
GP is the Shlomo and Michla Tomarin development chair, supported by the grants from GIF, Gruber foundation, IRG, ISF and Minerva. The work of LR was supported in part by  NSF grant PHY-0855591, NSF grant PHY-0556111 and the Fundamental Laws Initiative of the Harvard Center for the Fundamental
Laws of Nature.\\

\appendix

\section{One-loop dipole diagrams from vector-like quarks}\label{LoopApp}
We present here a detailed calculation of the one-loop diagrams giving rise to dipole operators in theories where the SM chiral spectrum ($Q^0,u^0,d^0$) is extended with heavy vector-like quarks ($Q^1+Q^1_-,u^1+u^1_-,d^1+d^1_-$), where the minus subscript denotes the wrong chirality fermions. We assume only one heavy SU(2)$_L$ doublet, one up-type singlet and one down-type singlet. Besides its simplicity such a setup completely describes the minimal RS (composite) framework with one level of KK states (strong resonances). The result is then straightforwardly generalized to RS models where each SM fermion comes along with a tower of vector-like KK states. As we are ultimately interested in flavor violating dipole operators it is enough to only consider Yukawa interactions whose Lagrangian is %
\begin{eqnarray}
-\mathcal{L}&\supset& y^{00}_u \bar{Q}^0\tilde H u^0+y^{00}_d \bar{Q}^0H d^0+y^{01}_u \bar{Q}^0\tilde H u^1+ y^{10}_u \bar{Q}^1\tilde H u^0+y^{11}_u \bar{Q}^1\tilde H u^1+{\rm h.c.}\nonumber\\
&&+y^{01}_d \bar{Q}^0 H d^1+ y^{10}_d \bar{Q}^1 H d^0+y^{11}_d \bar{Q}^1 H d^1+y^-_u \bar{Q}_-^1 \tilde{H}u_-^1+y^-_d \bar{Q}_-^1 Hd_-^1+{\rm h.c.}\nonumber\\
&&+m_Q \bar{Q}^1 Q_-^1+m_u\bar{u}_-^1u^1+m_d\bar{d}_-^1d^1+{\rm h.c.}\label{Leff}
\end{eqnarray}
where $H$ is the Higgs doublet, $\tilde H=i\sigma_2H^*$ and $m_{Q,u,d}\sim m_{\rm KK}\gg \langle H\rangle$. For the sake of definiteness we focus on one-loop diagrams matching to chromomatgnetic dipole operators of the form $\mathcal{O}_g=g_s\bar{Q}^0\tilde H \sigma_{\mu\nu}G^{\mu\nu}u^0$ where $g_s$ and $G_{\mu\nu}$ are the QCD gauge coupling and field strength respectively. We consider the transition $u^0(p)\to H+Q^0(p')+g(q)$, assuming the Higgs field carries no momentum. Lorentz symmetry and gauge invariance imply that the corresponding amplitude $\mathcal{M}$ can be decomposed as $\mathcal{M}=g_sT^a \epsilon_\mu^*(q)\mathcal{M}^\mu$ with
\beq
\mathcal{M}^\mu=A [\bar u(p') \gamma^\mu P_R u(p)]+B[i\bar{u}(p')\sigma^{\mu\nu}q_\nu P_R u(p)]\,,
\eeq
where $\epsilon_\mu(q)$ is the gluon polarization vector and $q=p-p'$. $A$ and $B$ are functions of $q^2$. The Wilson coefficient is then $C_g=-B/2$.
There are two types of diagrams contributing to the corresponding Wilson coefficient $C_g$, depending on whether the Higgs is attached to the internal or external quark line; we thus define $C_g=C_g^{\,\rm int}+C_g^{\,\rm ext}$. Notice that for both types it is sufficient to consider diagrams where the vector boson is radiated by an internal line, since diagrams with a vector boson attached to an external quark leg only contribute to the function $A$.

\subsection{1PI diagrams}
\begin{figure}[t]
\begin{center}
\includegraphics[width=0.48\textwidth]{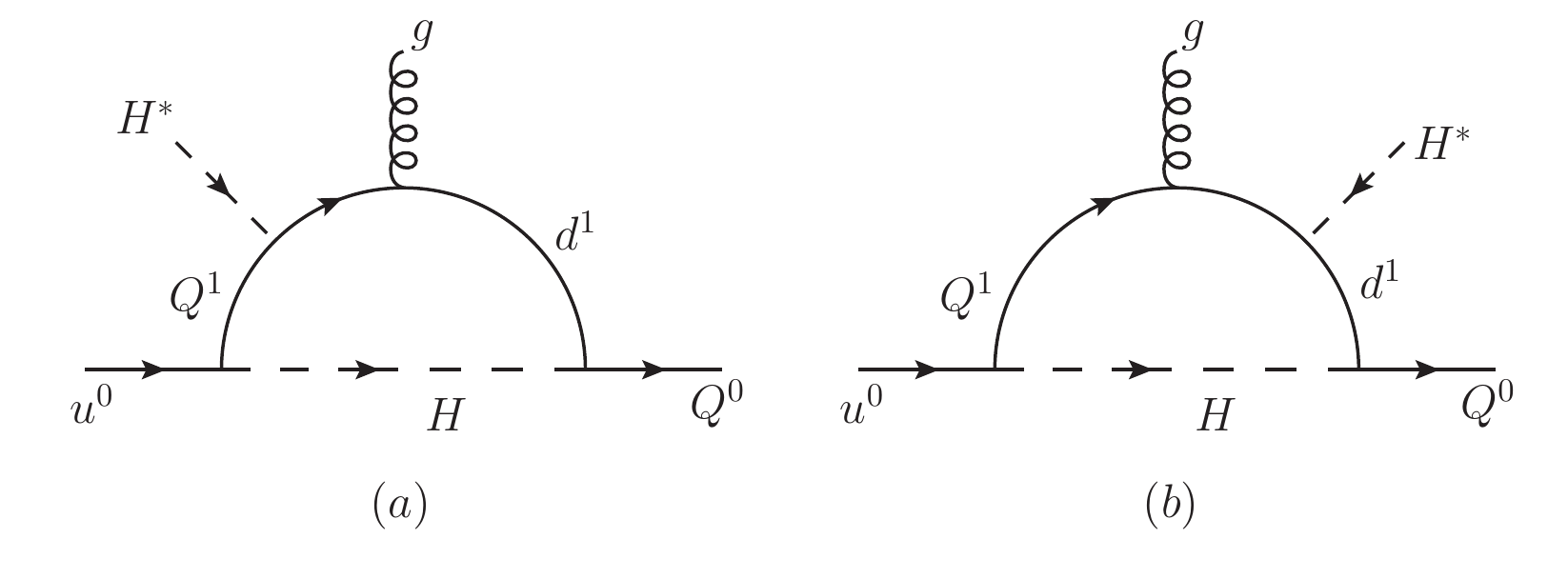}
\caption{{One-loop 1PI diagrams matching to chromomagnetic dipole operators.}}
\label{fig:EFT1PI}
\end{center}
\end{figure}

We consider first 1PI diagrams where the external Higgs is attached to the internal fermion line. The relevant diagrams are listed in Fig.\ref{fig:EFT1PI}. Notice there is no diagram with up-type quarks running in the loop thanks to a PQ-like symmetry
preserved by the Yukawa couplings and KK mass terms, under which the $\bar{Q}^0$ and the $\bar{Q}^1$ carry charge 1, whereas the Higgs field and the $Q^1_-$  carry charge $-1$. Such a  symmetry forces the up-type chiral states and heavy fermions of same chirality to be accompanied by the Higgs field and {\it not} its hermitian conjugate. Since there is only a single Higgs field, down-type Yukawa interactions break PQ and thus yield the only 1PI contributions.
The amplitudes corresponding to the diagrams in Fig.~\ref{fig:EFT1PI} are
\begin{eqnarray}
i\mathcal{M}_a&=& \int \frac{d^4k}{(2\pi)^4}\bar{u}(p')[iy_d^{01}P_R]\frac{i(\slashed k +\slashed p'+m_d)}{(k+p')^2-m_d^2}[iy_d^{11*}P_L
+iy_d^{-*}P_R]\nonumber\\
&&\times \frac{i(\slashed k +\slashed p'+m_Q)}{(k+p')^2-m_Q^2}[ig_sT^a\slashed \epsilon^*]\frac{i(\slashed k +\slashed p+m_Q)}{(k+p)^2-m_Q^2}[iy_u^{10}P_R]u(p)\frac{i}{k^2-m_H^2}\,,\label{Ma}\\
i\mathcal{M}_b&=& \int \frac{d^4k}{(2\pi)^4}\bar{u}(p')[iy_d^{01}P_R]\frac{i(\slashed k +\slashed p'+m_d)}{(k+p')^2-m_d^2}[ig_sT^a\slashed \epsilon^*]\frac{i(\slashed k +\slashed p'+m_d)}{(k+p')^2-m_d^2}\nonumber\\
&&\times[iy_d^{11*}P_L+iy_d^{-*}P_R]\frac{i(\slashed k +\slashed p+m_Q)}{(k+p)^2-m_Q^2}[iy_u^{10}P_R]u(p)\frac{i}{k^2-m_H^2}\,,\label{Mb}
%
%i\mathcal{M}_c&=&\int \frac{d^4k}{(2\pi)^4}\bar{u}(p')[iy_d^{01}P_R]\frac{i(\slashed k+m_d)}{k^2-m_d^2}[iy_d^{11*}P_L\nonumber\\
%&&+iy_d^{-*}P_R]\frac{i(\slashed k+m_Q)}{k^2-m_Q^2}[iy_u^{10}(i\sigma_2)P_R]u(p)[ieQ_H]\nonumber\\
%&&\times\frac{i^2\epsilon^*\cdot(2k-p-p')}{[(k-p)^2-m_H^2][(k-p')^2-m_H^2]}
\end{eqnarray}
where $P_{L(R)}$ is the left-handed (right-handed) chirality projector. %$\mathcal{M}_c$ only contributes to $C_F^{\,\rm int}$.
We kept the loop momentum integrals in four dimensions as we are only interested in finite dipole contribution.
After integrating over the loop momentum we find $32\pi^2C_g^{\,\rm int}=y_u^{01}y_d^{11*}y_d^{10}(I^{\,\rm int}_a+I^{\,\rm int}_b)+y_u^{01}y_d^{-*}y_d^{10}(J^{\,\rm int}_a+J^{\,\rm int}_b)$, where
\begin{eqnarray}
I^{\,\rm int}_a&=&\int dxdydz\bigg[\frac{1-3(x+y)}{\Delta}+\frac{zm_Q^2+q^2z(x+y)(1-x-y)}{\Delta^2}\bigg]\nonumber\\ %&&+\left(\frac{xm_d^2+q^2x(y+z)(1-y-z)}{\Delta_b^2}-\frac{3(y+z)-1}{\Delta_b}\right)\bigg]\,,\\
% &&+\delta_{VF}Q_H\left(\frac{3(y+z)-2}{\Delta_c}\right)\bigg]+\mathcal{O}(q^2)\,,\\
J^{\,\rm int}_a&=&m_Qm_d\int dxdydz\frac{(x+y+z)}{\Delta^2}\,,
%&&+\delta_{VF}Q_H\left(???\right)\bigg]+\mathcal{O}(q^2)\,,
\end{eqnarray}
with $\Delta=m_Q^2(y+z)+x m_d^2+m_H^2(1-x-y-z)-q^2z(x+y)$.
$I_b$ and $J_b$ are obtained from $I_a$ and $J_a$ with the replacements $\{x,m_d\}\leftrightarrow \{z,m_Q\}$.
The remaining Feynman integrals are over $0\leq x\leq 1$, $0\leq y\leq 1-x$ and $0\leq z\leq 1-x-y$. Integrating over the Feynman parameters and taking the limit of $m_H^2,q^2\ll m_{Q,d}^2$ yield
\beq
I^{\,\rm int}_a\simeq I^{\,\rm int}_b \simeq \mathcal{O}\left(\frac{m_H^2}{m_{Q,d}^4}\right)\,;\quad
J^{\,\rm int}_a\simeq\frac{r^3-2r\log r-r}{2m_d^2(r^2-1)^2}\,,\quad
J^{\,\rm int}_b\simeq\frac{r^{-1}+2r\log r-r}{2m_d^2(r^2-1)^2}\,,
\eeq
with $r\equiv m_Q/m_d$. We see that $I^{\,\rm int}_{a,b}$ vanish in the limit where $m_H\to 0$ (see also~\cite{AgashePetriello}). In that limit we thus find
\beq
C_g^{\rm int} = \frac{1}{64\pi^2}\frac{y_u^{01}y_d^{-*}y_d^{10}}{m_Qm_u}+\mathcal{O}\left(\frac{m_H^2}{m_{Q,d}^4}\right)\,.
\eeq

\subsection{Non-1PI diagrams}
\begin{figure}[t]
\begin{center}
\includegraphics[width=0.7\textwidth]{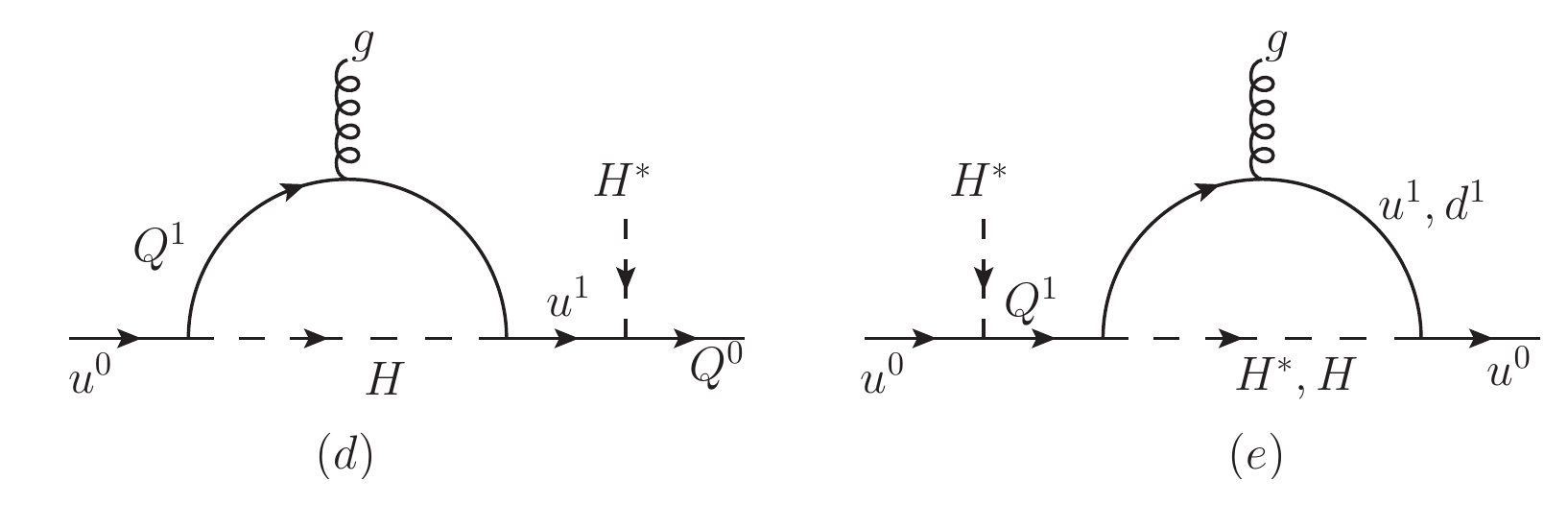}\\
\caption{{One-loop non-1PI diagrams matching to chromomagnetic dipole operators.}}
\label{fig:EFTnon1PI}
\end{center}
\end{figure}

There are additional non-1PI diagrams contributing to the operator $\mathcal{O}_g$ below the heavy scale $m_{\rm KK}$. They correspond to diagrams with a Higgs vertex on the external quark lines and an intermediate heavy quark propagator, as shown in Fig.\ref{fig:EFTnon1PI}. The corresponding amplitudes are
\begin{eqnarray}
i\mathcal{M}_e&=& \int \frac{d^4k}{(2\pi)^4}\bar{u}(p')[iy_u^{01}P_R]\frac{i(\slashed k+ \slashed p'+m_{u,d})}{(k+p^{\prime })^2-m_{u,d}^2}[ig_sT^a\slashed \epsilon^*]\frac{i(\slashed k +\slashed p+m_{u,d})}{(k+p)^2-m_{u,d}^2}\nonumber\\
&&\times[iy_{u,d}^{11*}P_L+iy_{u,d}^{-*}P_R]\frac{i(\slashed p+m_Q)}{p^2-m_Q^2}
[iy_u^{10}P_R]u(p)\frac{i}{k^2-m_H^2}\,,\\
i\mathcal{M}_f&=&\int \frac{d^4k}{(2\pi)^4}\bar{u}(p')[iy_u^{01}P_R]\frac{i(\slashed p'+m_u)}{p^{\prime 2}-m_u^2}[iy_u^{11*}P_L+iy_u^{-*}P_R]\frac{i(\slashed k +\slashed p'+m_Q)}{(k+p')^2-m_Q^2}\nonumber\\
&&\times[ig_sT^a\slashed \epsilon^*]\frac{i(\slashed k +\slashed p+m_Q)}{(k+p)^2-m_Q^2}[iy_u^{10}P_R]u(p)\frac{i}{k^2-m_H^2}\,.
\end{eqnarray}
Notice that contributions $\propto y_u^{01}y_{u,d}^{11*}y_{u,d}^{10}$ vanish when the external quark leg is on-shell because the corresponding diagrams are proportional to the external quark equation of motion $\slashed p u(p)$ or $\bar{u}(p')\slashed p'$. Hence the remaining non-1PI contribution to the Wilson coefficient is
$32\pi^2C_g^{\,\rm ext}=y_u^{01}y_u^{-*}y_u^{10}(J^{\,\rm ext}_{u,e}+J^{\,\rm ext}_{u,f})+y_u^{01}y_d^{-*}y_d^{10}J_{d,e}^{\,\rm ext}$,
where
\beq
J^{\,\rm ext}_{u,e}=\frac{m_u}{m_Q}\int dxdy\frac{(x+y)}{\Delta_u}\,,\quad
J^{\,\rm ext}_{u,f}=\frac{m_Q}{m_u}\int dxdy\frac{(x+y)}{\Delta_Q}\,,
\eeq
and $J^{\,\rm ext}_{d,e}=J^{\,\rm ext}_{u,e}(u\to d)$ with $\Delta_X=m_X^2(x+y)+m_H^2(1-x-y)-xyq^2$, $X=Q,u,d$.
Integrating over the Feynman parameters and taking the limit $m_h^2,q^2\ll m_{Q,u,d}^2$, we find
\beq
J_{u,e}^{\, \rm ext}\simeq J_{u,f}^{\, \rm ext}\simeq  \frac{1}{2m_Qm_u}\,,\quad
J_{d,e}^{\, \rm ext}\simeq \frac{1}{2m_Qm_d}\,.
\eeq
%
%The explicit calculation presented above shows that no (chromo)magnetic dipole operators are generated at the dimension 6 level in a limit where the Yukawas coupling the wrong chirality states are absent, {\it i.e.} $y_{u,d}^-=0$. Although there does not seem to be any symmetry underlying this result, it still suggests that dipole operators are only induced through operators of dimension 8 or higher.\\
%Therefore dipole operators arising from neutral and charged Higgs one-loop diagrams are UV finite in the limit of RS models where the Higgs is an IR localized field.
The non 1PI diagrams thus yield the following contribution to the chromomagnetic dipole operators
\beq
C^{\rm ext}_g=\frac{1}{32\pi^2}\frac{y_u^{01}y_u^{-*}y_u^{10}}{m_Qm_u}+\frac{1}{64\pi^2}\frac{y_u^{01}y_d^{-*}y_d^{10}}{m_Qm_d}\,.
\eeq

\subsection{Matching to RS models}

We finally comment on RS models, which contain several towers of vector-like KK states: $Q^n$, $u^n$ and $d^n$ of masses $m_{Q,u,d}^{(n)}\simeq n m_{\rm KK}$. The above result is easily generalizable to those models. Upon resuming the doublet and singlet KK towers the Wilson coefficient is $C_g=\sum_{n,m}C_g^{(n,m)}$ where
\beq
C_g^{(n,m)}= \frac{y_u^{0n}y_d^{nm*}y_d^{m0}}{32\pi^2m_Q^{(n)}m_d^{(m)}}+ \frac{y_u^{0n}y_u^{nm*}y_u^{m0}}{32\pi^2m_Q^{(n)}m_u^{(m)}}\,.
\eeq
Since the two terms above involve {\it a priori}  independent phases they cannot be combined. We chose to focus on the $y_u^3$ contribution in the main text since flavor anarchy is only assumed in the up-sector.

\subsection{Electromagnetic dipole contributions}

\begin{figure}[t]
\begin{center}
\includegraphics[width=0.30\textwidth]{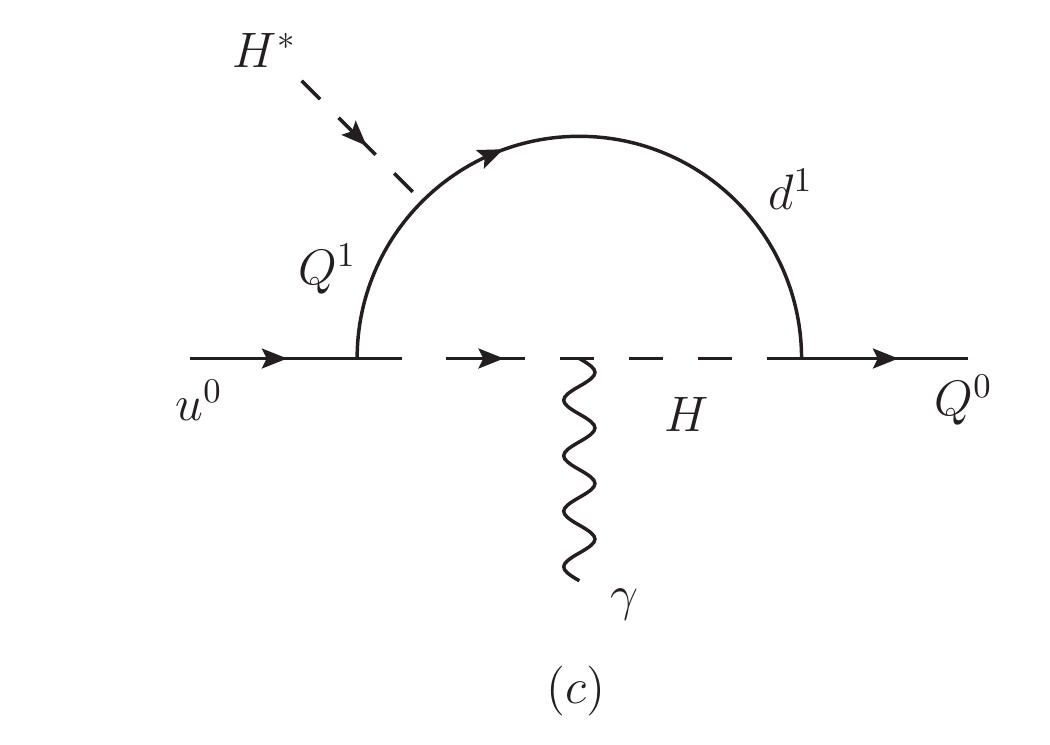}
\caption{{Additional one-loop 1PI diagram inducing magnetic dipole operators.}}
\label{fig:EFTphoton}
\end{center}
\end{figure}

When considering QED magnetic dipole operators like $\mathcal{O}_\gamma = e\bar{Q}^0\tilde H\sigma_{\mu\nu}F^{\mu\nu}u^0$ there is another 1PI diagram where the photon is radiated by the (charged) Higgs line, as in Fig.~\ref{fig:EFTphoton}. The corresponding amplitude is
\begin{eqnarray}
i\mathcal{M}_c&=&\int \frac{d^4k}{(2\pi)^4}\bar{u}(p')[iy_d^{01}P_R]\frac{i(\slashed k+m_d)}{k^2-m_d^2}[iy_d^{11*}P_L+iy_d^{-*}P_R]\frac{i(\slashed k+m_Q)}{k^2-m_Q^2}\nonumber\\
&&\times[iy_u^{10}P_R]u(p)[ieQ_H]\frac{i^2\epsilon^*\cdot(2k-p-p')}{[(k-p)^2-m_H^2][(k-p')^2-m_H^2]}\,,
\end{eqnarray}
which yields a contribution to the $\mathcal{O}_\gamma$ Wilson coefficient of $32\pi^2C_\gamma^{\,\rm int}\supset y_u^{01}y_d^{11*}y_d^{10}I_c^{\,\rm int}+y_u^{01}y_d^{-*}y_d^{10}J_c^{\,\rm int}$
where
\beq
I_c^{\,\rm int}=\int dxdydz\left[\frac{3(y+z)-2}{\Delta_\gamma}-\frac{q^2yz(1-y-z)}{\Delta_\gamma^2}\right]\,,\quad
J_c^{\,\rm int}=m_Qm_d\int dxdydz\frac{(y+z-1)}{\Delta_\gamma^2}\,,
\eeq
with $\Delta_\gamma=m_d^2(1-x-y-z)+xm_Q^2+m_H^2(y+z)-q^2yz$. After Feynman parameter integration we find in the limit $m_H^2,q^2\ll m_{Q,d}^2$
\beq
I_c^{\,\rm int}\simeq \mathcal{O}\left(\frac{m_H^2}{m_{Q,d}^4}\right)\,,\quad J_c^{\,\rm int}\simeq \frac{1}{2m_Qm_d}\,.
\eeq
Here again, $I_c^{\,\rm int}$ vanishes in the $m_H\to 0$ limit. Combining the amplitude from the diagram in Fig.\ref{fig:EFTphoton} with those of the diagrams in Figs.~\ref{fig:EFT1PI}-\ref{fig:EFTnon1PI}, where the gluon is replaced by a photon, yields
\beq
C_\gamma=\frac{1}{192\pi^2}\frac{y_u^{01}y_d^{-*}y_d^{10}}{m_Qm_d}+\frac{1}{48\pi^2}\frac{y_u^{01}y_u^{-*}y_u^{10}}{m_Qm_u}\,.
\eeq

%Furthermore the PQ symmetry will ensure that the neutral Higgs diagrams does not receive UV sensitive contributions.\\
%

\section{A toy flat extra dimension model}\label{flatXD}
We prove here why the overlap function $\mathcal{O}_\beta$ defined in Eq.~\eqref{Obdef} is finite and does not depend on the Higgs width when the latter is quasi-localized on the IR.  We show this result in the simple case of  a flat extra dimension. Nonetheless we stress again that the flat case should provide a reasonable approximation to the warped case as far as the width of the Higgs profile is small so that deviations from the flat geometry are subdominant.
Let us consider a flat extra dimension of size $L$ where the wrong chirality modes are given by a set of sine functions, whereas the right modes correspond to towers of cosine functions.

The Higgs profile is taken to be peaked towards the IR like $e^{-z/\epsilon L}$, where $z$ is the extra dimension coordinate. We focus on the contribution to the dipole operator in the limit where  $\epsilon\ll1$.
Since the cosine wave-functions of the right chirality modes are approximately flat in the region where the Higgs profile is peaking,  their couplings to the zero mode have very weak dependence on the Higgs profile in the small $\epsilon$ limit. The Higgs overlap function $\mathcal{O}_\beta$ is then proportional to
\beq
\epsilon^{-1}\mathcal{O}_\beta^{\rm flat}\propto \epsilon^{-1}\sum_{n,m=1}^{\infty}{1\over m n}\int _0^L dz
\sin \left(n \pi z/L\right)\sin \left(m \pi z/L\right)  e^{-z/\epsilon L}\,,
\eeq
where we divide both sides by $\epsilon$ in order to maintain a finite value for the Yukawa overlap integral, equivalent to rescaling the 5D Yukawa by $\sqrt{1+\beta}$ for the warped case with the polynomial Higgs profile in Eq.~\eqref{Hprofile}.
For a high enough KK level, then, one can replace the sum over $n$ and $m$ by a continuous integral. Moving to radial coordinates, defining $r^2=n^2+m^2$ and $\tan \theta=n/m$, the resulting integral can be solved analytically.
To understand why the sum is finite and independent of $\epsilon$ let us first integrate  over $z$ and $\theta$, leaving an $r$-dependent integral of (to leading order in $\epsilon$):
\beq\label{flatint}
\epsilon^{-1}\mathcal{O}_\beta^{\rm flat} \propto \int_{\sqrt2}^\infty r dr \frac{(\epsilon\pi)^2}{(1 + \epsilon^2 \pi^2 r^2) \sqrt{
 1 + 2 \epsilon^2 \pi^2 r^2}}\,.
\eeq
The change of variable $x=\epsilon \pi r$ explicitly shows that the result is indeed $\epsilon$-independent as anticipated.
Furthermore one can see from Eq.~\eqref{flatint} that the dominant contribution to the KK sums comes from heavy modes with $r\sim 1/\epsilon$, however the width of the distribution is such that $1/\epsilon^2$ modes contribute to the integral, yielding an $\epsilon$ independent result.
Eq.~\eqref{flatint} also shows that for any small but finite $\epsilon$  the ultra heavy modes with $r\gg 1/\epsilon$ give a negligible contribution to the integral, because such modes rapidly oscillate within the Higgs profile's width, leading to a single KK sum and thus a highly suppressed contribution. This fact also explains why the total dipole contribution is guaranteed to be finite.

\end{document}